\documentclass[twocolumn,prl,showpacs]{revtex4}

\begin{document}

\topmargin-1.0cm

\title {Degeneracy in Density Functional Theory: Topology in $v$- and $n$-Space}
\author {C. A. Ullrich}
\altaffiliation[Present Address: ]
{Department of Physics, University of Missouri-Rolla, Rolla, MO 65409}
\author{W. Kohn}
\affiliation
{Department of Physics,  University of California, Santa Barbara, CA 93106}
\date{\today}
\begin{abstract}
This paper clarifies the topology of the mapping between $v$- and $n$-space in fermionic systems.
Density manifolds corresponding
to degeneracies $g=1$ and $g>1$ are shown to have the same mathematical measure: 
every density near a $g$-ensemble-$v$-representable
($g$-VR) $n({\bf r})$ is also $g$-VR (except ``boundary densities'' of lower measure).
The role of symmetry and the connection between 
$T=0$ and $T=0^+$ are discussed. A lattice model and the Be-series are used as illustrations.
\end{abstract}
\pacs{31.15.Ew, 31.15.Ar, 71.10.Ca, 71.15.Mb}
\maketitle
Density functional theory (DFT) of 
electrons in an external potential $v({\bf r})$ uses the ground-state
density $n({\bf r})$ as basic variable \cite{HK,KS}.
$n({\bf r})$ uniquely determines
$v({\bf r})$ (apart from a constant);
conversely, for a given number, $N$, of electrons, 
$v({\bf r})$ obviously determines $n({\bf r})$ uniquely if and only if the ground
state is non-degenerate. These facts point to the interesting
role of degeneracy in the mapping $v({\bf r})\!\Longleftrightarrow n({\bf r})$.

In 1982-3, Levy \cite{Levy} and Lieb \cite{Lieb} 
discovered examples of well-behaved density functions, 
$n({\bf r})$, which {\em could not} be reproduced as non-degenerate, 
non-interacting ground-state densities of {\em any}
$v({\bf r})$, and were called non-$v$-representable (non-VR).
In these examples, the  $n({\bf r})$ {\em could} be
reproduced as weighted averages of densities of degenerate
ground states corresponding to a $v({\bf r})$.

Soon afterwards, Kohn \cite{Kohn} adopted a lattice version of the
Schr\"odinger equation, ${\bf r} \to r_l$ ($l=1,\ldots,M$), in which 
$v({\bf r})$ and $n({\bf r})$ become
$v_l$ and $n_l$ and can be viewed as $M'$-dimensional ($M'$-D) vectors:
$\vec{v} \equiv (v_1,\ldots,v_{M'})$, with $M' \equiv M-1$, 
$v_M = 0$; and
$\vec{n} \equiv (n_1, \ldots, n_{M'})$, with $n_M = N - (n_1 +\ldots
+ n_{M'})$ \cite{footnote}.
The $M$ lattice points are enclosed by boundary points on which
$\vec{v} = +\infty$ and, accordingly, all wave functions vanish.
It was shown in \cite{Kohn} that in the $M'$-D $\vec{n}$-space 
there are finite $M'$-D regions in
which all ``points'' (densities) are VR.

In 1985, Chayes {\em et al.} (CCR) \cite{Chayes}, using a similar
lattice model, proved the following important result: Any well-behaved
$\vec{n}$ ($0 < n_l \le 1$) can be uniquely represented 
as weighted average of $g$ 
degenerate ground-state densities associated with a $\vec{v}$.
We call the special case $g=1$ 1-VR,
and the general case ($g\ge 1$) $g$-VR.
Thus, $\vec{n}$-space is filled by manifolds $S^g$,
in each of which $\vec{n}$ is $g$-VR.
Each $S^g$ maps on a corresponding manifold $Q^g$ in $\vec{v}$-space.

This paper aims at
an understanding of the topologies of the regions $S^g$ and $Q^g$ 
and of the nature of the boundaries between them. This has mathematical and 
physical significance, e.g. in the search for
self-consistent solutions of Kohn-Sham (KS) equations \cite{ullrichkohn}.
We shall see that $g$-VR densities with $g>1$ are {\em not} mathematically 
exceptional and, in particular, do {\em not} depend on symmetries of the
potential.  As illustrations we shall
present a finite lattice model and the Be-series for $Z\to \infty$ \cite{schipper}.
We also discuss the relationship between $T=0$ ensembles and $T=0^+$ thermal
ensembles. We limit ourselves to systems that are everywhere nonmagnetic.

{\bf Interior regions in $\vec{\lowercase{n}}$-space}.
We first recapitulate Ref. \cite{Kohn} for 1-VR densities, and
then generalize to $g$-VR.

a) $g=1$.
Since there is a non-degenerate non-interacting ground state with a finite gap,
corresponding to some $\vec{v}$, we can use non-degenerate perturbation theory 
to calculate the first-order density change, $^1n_l$, due to a weak perturbing
potential $^1v_l$:
\begin{equation} \label{dndv}
^1n_l = \sum_{l' =1}^{M'} \chi_{ll'} \:{^1v_{l'}} \;,
\qquad l=1,\ldots,M' \;.
\end{equation}
Since, according to \cite{HK}, the density $\vec{n}$ of a 
non-degenerate ground state uniquely determines $\vec{v}$
(no arbitrary constant since $v_M=0$), 
the homogeneous set of equations corresponding
to (\ref{dndv}) has no solution other than $^1v_{l'} \equiv 0$, so that (\ref{dndv})
can be inverted:
\begin{equation} \label{dvdn}
^1v_l = \sum_{l'=1}^{M'} \chi_{ll'}^{-1} \: {^1n_{l'}} 
\qquad l=1,\ldots,M' \;.
\end{equation}
Thus, any first-order change of $\vec{n}$ preserving $N$ produces a density which is also
1-VR. This is the case in the entire $M'$-D neighborhood
surrounding the point $\vec{n}$, in which $\Delta$ remains positive.

b) $g>1$ \cite{wigner}.
We denote by $\Psi_1, \ldots, \Psi_g$ a set of $g$
orthogonal degenerate ground-state wave functions corresponding to 
$\vec{v}^g$, with common energy $E$ and finite gap $\Delta$. The 
remaining eigenfunctions are  $\Psi_{g+1},\ldots$.
What conditions must be imposed on infinitesimal [${\cal O}(\lambda)$] potential changes
$^1\vec{v}$ which will preserve this degeneracy? 
In the space of the $\Psi_j$, the perturbed Hamiltonian matrix is
\begin{equation}\label{proof1}
H_{ij} = E_i \delta_{ij} +  {^1V}_{ij}\; \qquad i,j=1,2,\ldots \;,
\end{equation}
$E_i=E$ ($i=1,\ldots,g$), $E_i \ge E+\Delta$
($i\ge g+1$), and $^1V_{ij} \equiv \sum_{l=1}^{M'} \langle i | {^1v_l}|j\rangle$.
Because of the finite $\Delta$, the off-diagonal
$^1V_{ij}$ with $i\le g$ and $j\ge g+1$, or vice versa, can be removed to all
orders in $\lambda$ by orthogonal transformations leading to the decoupled
$g\times g$ block Hamiltonian
\begin{equation}\label{proof2}
\tilde{H}_{ij} = E \delta_{ij} +  {^1V}_{ij} + {\cal O}(\lambda^2)\;
\qquad i,j=1,\ldots,g \;.
\end{equation}
Diagonalization by an orthogonal transformation $T_{ij} = \delta_{ij} +
{^1 t}_{ij} +  {\cal O}(\lambda^2)$ and the requirement that the eigenvalues
remain degenerate gives a transformed Hamiltonian 
\begin{equation}\label{proof3}
\bar{H}_{ij} \equiv \left( T^{-1} \tilde{H} T\right)_{ij}
= (E + {^1\epsilon})\delta_{ij} + {\cal O}(\lambda^2) \;.
\end{equation}
On inverting the transformation and noting that every matrix commutes with 
$\delta_{ij}$, one finds
$\tilde{H}_{ij} =\bar{H}_{ij}+ {\cal O}(\lambda^2)$.
Comparison with Eq. (\ref{proof2}) leads to $\frac{1}{2} (g-1)(g+2)$
conditions of equal diagonal and vanishing off-diagonal $^1V_{ij}$.
Thus, every point $\vec{v}^g$ with a 
$g$-fold degenerate ground state and finite $\Delta$ is embedded in a 
manifold in $\vec{v}$-space of dimension $D^g=\left[ M' -\frac{1}{2} (g-1)(g+2)\right]$, in which
the degeneracy $g$ and a finite $\Delta$ are preserved. Ground-state
degeneracies in $\vec{v}$-space are ``rare'' in the above sense.

Each $\vec{v}^g$ gives rise to a  set of ensemble densities,
\begin{equation} \label{nn}
\vec{n} = \sum_{j=1}^g w_j \vec{n}_j\;, \qquad
0 < w_j \le 1, \qquad \sum_{j=1}^g w_j = 1 \;,
\end{equation}
with $\vec{n}_j = \vec{n}_j\left(R^g \Psi_j\right)$, 
where $R^g$ is a $g$-D orthogonal transformation [$\frac{1}{2}g(g-1)$ parameters]
and the $w_j$ are normalized weights ($g-1$ parameters). Thus,
all $\vec{n}'$ in a finite $M'$-D neighborhood
enclosing a $g$-VR $\vec{n}$, defined by $\Delta(\vec{n}')>0$ and $w_j>0$, are also $g$-VR:
Degeneracy in $\vec{n}$-space is {\em not} ``rare''.

{\bf Boundary surfaces in $\vec{\lowercase{n}}$-space}.
Except for boundaries of ${\rm D}\le M'-1$,
$\vec{n}$-space is completely filled by
$M'$-D regions $S^g$, with degenerate ground-state levels and a finite positive
gap to the nearest excited state.

\begin{figure}
\unitlength1cm
\begin{picture}(5.0,3.5)
\put(-5.,-8.8){\makebox(5.0,3.5){
\includegraphics{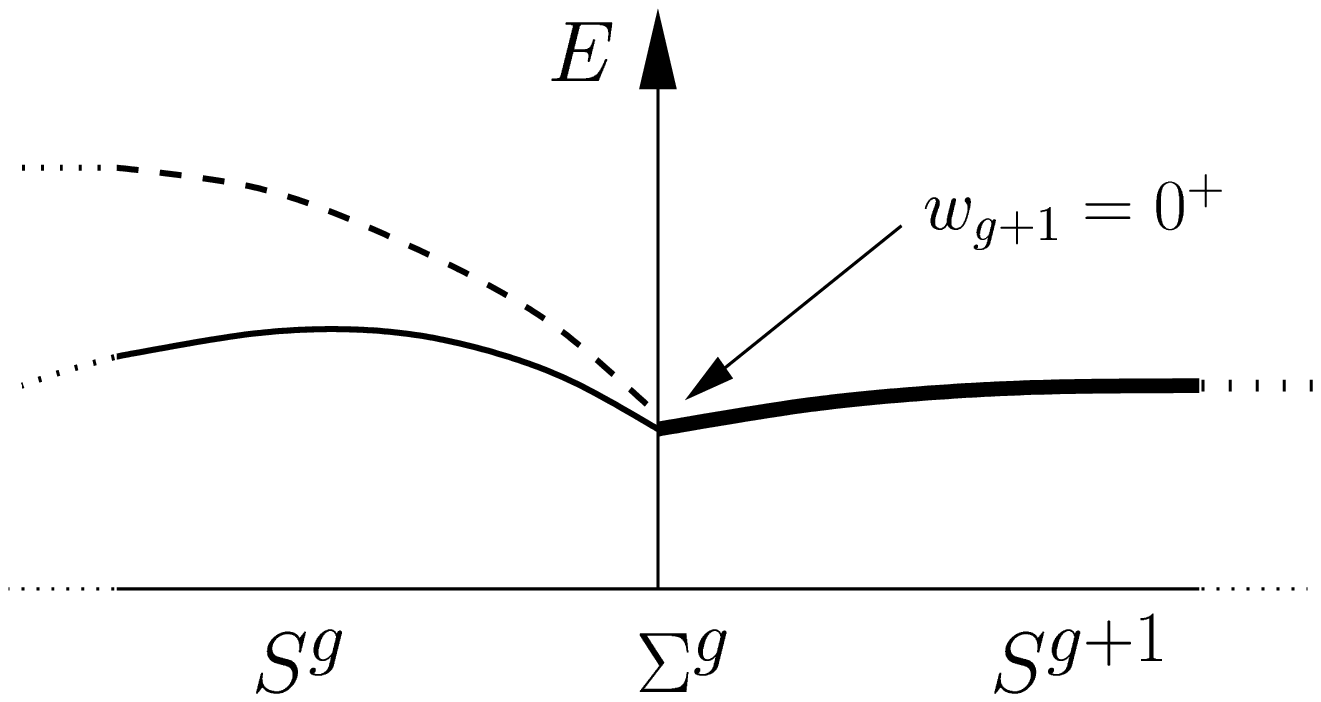}}}
\end{picture}
\caption{
Two interior regions in the $M'$-D $\vec{n}$-space,
$S^g$ and $S^{g+1}$, separated by a $(M'-1)$-D boundary $\Sigma^g$. 
}
\label{figure1}
\end{figure}

Apart from ``corners'' (${\rm D}\le M'-2$), interior regions $S^g$ in
$\vec{n}$-space are bounded by $(M'-1)$-D internal and external ``surfaces''
$\Sigma^g$, see Fig. \ref{figure1}. Each
internal $\Sigma^g$ separates two interior regions $S^g$ and $S^{g+1}$. A
point $\vec{n}$ {\em on} such a $\Sigma^g$ corresponds to a $(g+1)$-fold
degenerate ground state but with one of the $g+1$ weights equal to zero.
On the $g$-side of $\Sigma^g$ there is a $g$-fold degeneracy and a small
gap to the $(g+1)$st state opens up from $0$. On the $(g+1)$-side there is
a $(g+1)$-fold degeneracy and one of the $g+1$ weights starts from 
$0$ and becomes positive. 

External boundaries correspond to either one of the $n_l$ becoming zero,
reflecting particle conservation; or one of the $n_l$ becoming $1$, 
reflecting the Pauli principle. For examples of both kinds, 
see the lattice model below. 


{\bf Imposed symmetry conditions}.
Consider a Hamiltonian, invariant under a group
$G$ of the lattice with irreducible representations $h$ with ${\rm D}=g_h$.
Let the ground state have total degeneracy $g = \sum_h m_h g_h$
due to $m_h$ occurrences of $h$.
If a small perturbation $H'$ {\em respecting the group $G$} 
is imposed, how many further conditions must be met by  $H'$ so that
the degeneracy is maintained?

We have shown above that the perturbation must have vanishing off-diagonal
matrix elements (MEs) between all degenerate ground states, and equal diagonal
MEs.

{\em Off-diagonal MEs}. The symmetry of $H'$ assures immediately
that all off-diagonal MEs between $h \ne h'$ vanish \cite{tinkham}. Off-diagonal
MEs between two states of the same $h$ involve, by the
Wigner-Eckart Theorem, only one independent parameter. The total number of 
independent parameters characterizing all off-diagonal MEs
between the degenerate ground states is therefore 
$\sum_h m_h (m_h-1)/2$.

{\em Diagonal MEs}. The requirement that all diagonal MEs
of $H'$ are equal is automatically satisfied
for states belonging to the same $h$, and involves $\sum_h m_h - 1$ conditions
for states belonging to different $h$.

Thus, to ensure that the $g$-fold degenerate ground state maintains its full
degeneracy requires the imposition of $\sum_h [m_h (m_h-1)/2 + m_h] - 1$
conditions.

Next we ask: for a  ground state with $g = \sum_h m_h g_h$, how many 
free parameters are there to create ensemble densities
invariant under group $G$ of the Hamiltonian?

The most general ensemble density obtainable from 
$g$ degenerate ground states $\Psi_1,  \ldots, \Psi_g$ is
$\sum_{\mu=1}^g w_\mu \: n(\Psi_\mu')$, $\Psi_\mu' = R^g \Psi_\mu$.
However, the $\Psi_\mu'$ must be
chosen in a restricted way to assure that the density has the 
full symmetry of the Hamiltonian: Let $\varphi_k^{h,\nu_h}$
($k=1,\ldots,g_h$) be a set of orthogonal eigenfunctions, constituting 
a basis for a given $h$ of the group $G$, with occurrence
$\nu_h=1,\ldots,m_h$, such that  all 
$\varphi_k^{h,\nu_h}$  transform in the same way under 
$G$. The most general density of the required symmetry is 
\begin{equation}
\vec{n} = \sum_{h,\nu_h} w^{h,\nu_h} \sum_k 
\vec{n}(R^{\nu_h}\varphi_k^{h,\nu_h}) \;.
\end{equation}
The total number of free parameters is given by those
in the rotation $R^{\nu_h}$ plus the number of independent 
weights $w^{h,\nu_h}$, i.e. $\sum_h [m_h (m_h-1)/2 + m_h] - 1$.
We observe that the total number of constraints on $H'$ is equal to the 
total number of free parameters in $\vec{n}$. Therefore the
dimensionality of the density invariant under $G$ associated with any set
of degenerate states is the same, independent of $G$.

\begin{figure}
\unitlength1cm
\begin{picture}(5.0,3.5)
\put(-4.5,-7.6){\makebox(5.0,3.5){
\includegraphics{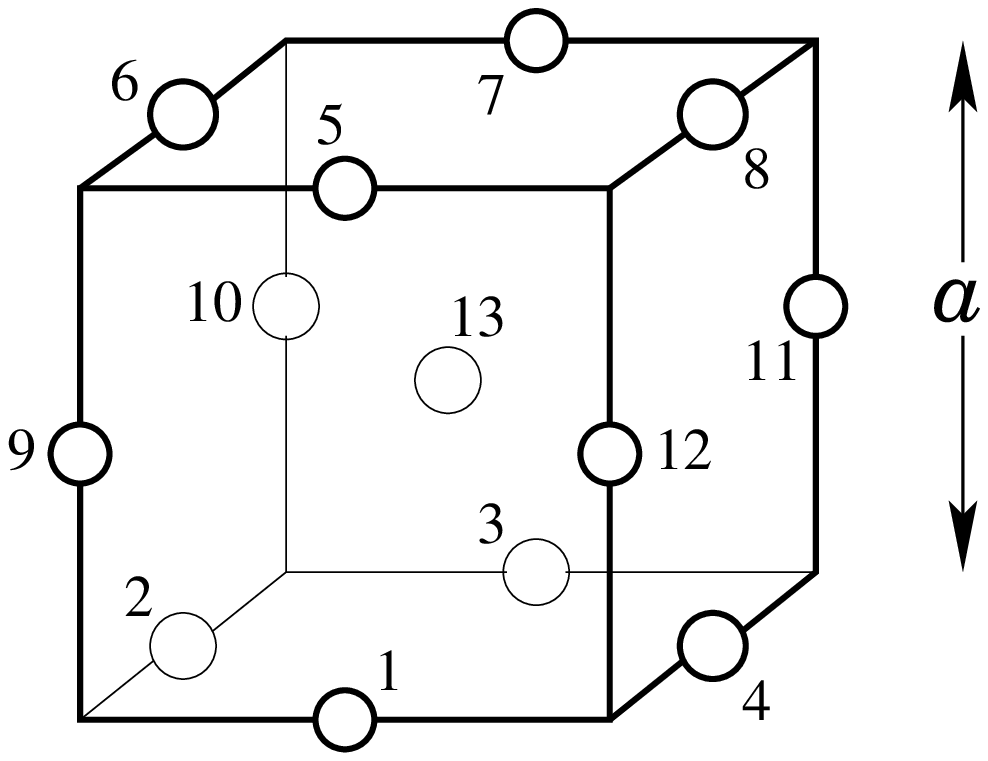}}}
\end{picture}
\caption{
13-point fcc lattice with $C_{4h}$ symmetry: $v_l=V_1$ ($V_2$)
on the sites $l=\mbox{1--8}$ (9--12), and $v_{13}=0$.}
\label{figure2}
\end{figure}

{\bf Lattice Example}.
We illustrate our topological results in an example 
including symmetry requirements in $\vec{v}$- and $\vec{n}$-space.
Consider the singlet ground state of 4 non-interacting electrons, 
or, equivalently, 2 spinless non-interacting fermions, whose wave functions $\varphi_l^j$
are defined on a 13-point fcc lattice (see Fig. \ref{figure2}),
and vanish on all exterior points. 
The discrete Schr\"odinger equation is
\begin{equation}\label{eq1}
\left[ - \frac{\hbar^2}{2m} \nabla_l^2 + v_l \right] \varphi_l^j        
= \varepsilon^j \varphi_l^j \;, \quad j,l=1,\ldots,13\;,
\end{equation}
where, for fcc lattices, 
$ \nabla_l^2 f_l = [\sum_k f_k - 12 f_l]/a^2$, with $k$ running over the
nearest neighbors of lattice site $l$. Energies 
and potentials will be measured in units of $2m a^2/\hbar^2$.

We consider cases of $C_{4h}$ symmetry:
$v_l \equiv V_1$, $l=\mbox{1--8}$; $v_l \equiv V_2$,
$l=\mbox{9--12}$; $v_{13}=0$. 
The lattice wave functions belong to the following
representations of $C_{4h}$ \cite{tinkham}:
$A_g$ ($s$-like), $E_u$ and $A_u$ ($p$-like), 
$E_g$ and  $B_g$ ($d$-like), and $B_u$ ($f$-like). $A_g$ occurs 
three times ($1A_g$, $2A_g$, $3A_g$), 
$E_u$ occurs twice ($1E_u$, $2E_u$), 
all other representations occur only once. The eigenfunctions and energies
required solutions of 3-, 2- and 1-D secular equations.

\begin{figure}
\unitlength1cm
\begin{picture}(5.0,13.3)
\put(-7.5,-12.4){\makebox(5.0,13.3){
\includegraphics{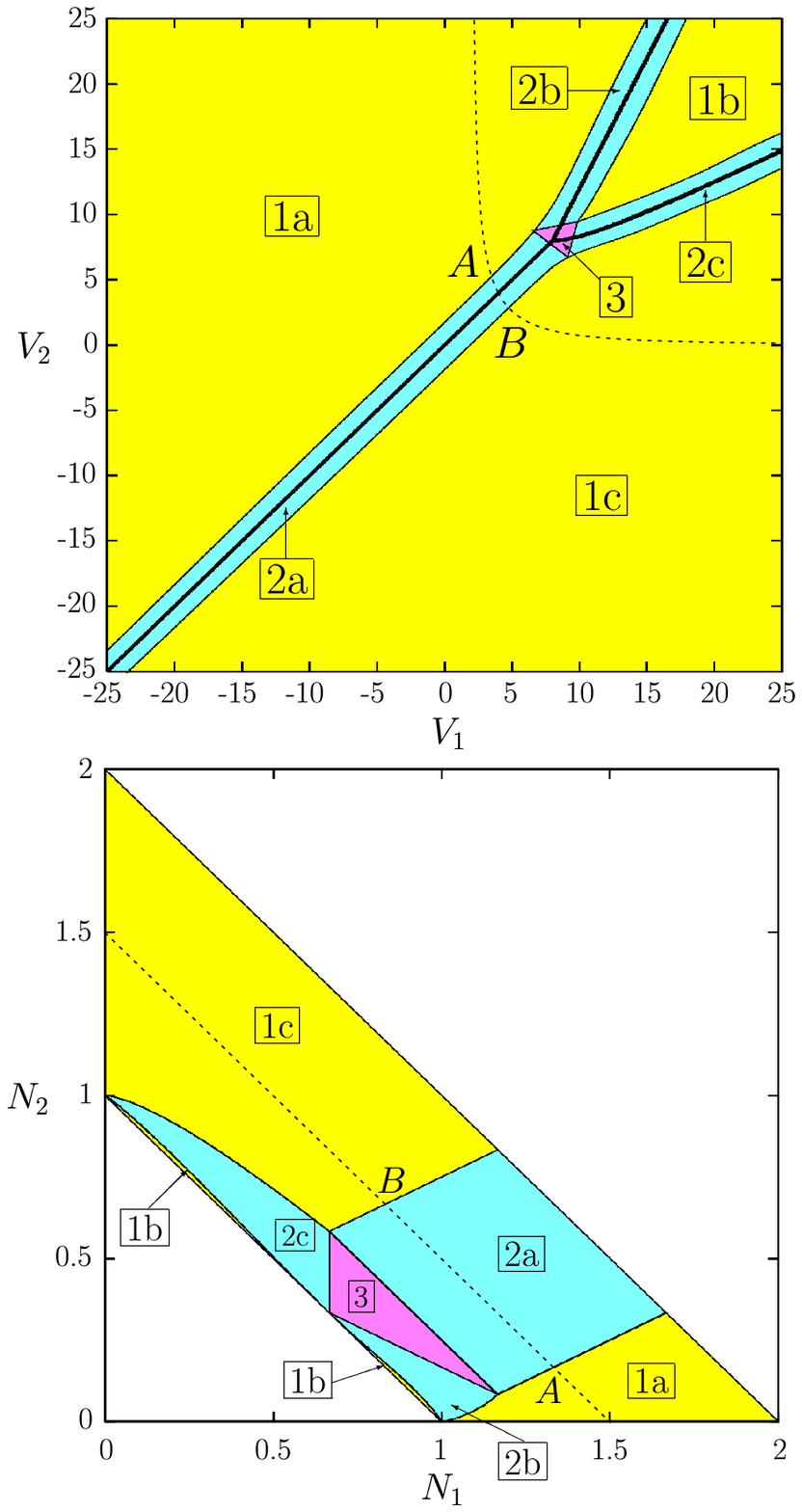}}}
\end{picture}
\caption{Top: ground states of 2 non-interacting spinless
fermions on the lattice of Fig. \ref{figure2}, with potentials of $C_{4h}$ symmetry.
Bottom: associated region of allowed densities.
}
\label{figure3}
\end{figure}

Consider the ground state of 2 non-interacting spinless fermions.
The lowest level is
always $1A_g$ ($1s$-like). The next level depends on 
$V_1$ and $V_2$, see Fig. \ref{figure3}.
The $(V_1,V_2)$ plane divides into three distinct, infinitely extended
regions (shown in yellow), in each of which the 
eigenfunction of the second level belongs to a particular irreducible 
representation of $C_{4h}$: Region 1a, $1A_u$  
($2p_z$-like); region 1b, $2A_g$ ($2s$-like); and region 1c, 
$1E_u$ ($2p_x$,$2p_y$-like). {\em Inside} each region, the
ground state is non-degenerate (not counting degeneracies dictated by symmetry,
such as in the 2-D representation $1E_u$).

On the boundary lines 2a, 2b and 2c (shown in blue), 
crossings of levels belonging to different representations of 
$C_{4h}$ occur, leading to 
two-fold (accidental) degeneracies. Along 2a, the 
potential has cubic symmetry
($V_1=V_2$), and the two lowest $p$-like levels ($1A_u$, $1E_u$) 
coincide. On 2b, the $2A_g$ and $1A_u$ levels cross, and on 2c, 
the $2A_g$ and $1E_u$ levels cross.
All three boundary lines meet in a single point, ``3'', at 
$V_1=V_2=8$. At this point only, the ground state has a three-fold accidental
degeneracy. Note that the degeneracies on
2b and 2c are not due to additional potential symmetries.

As shown above, a ground
state with restricted symmetry maintains its degeneracy if $\sum_h [m_h(m_h-1)/2
+m_h]-1$ conditions are met. In the present example, $m_h=1$ always. Regions
2a, 2b, 2c are therefore lines (1 condition), region 3 is a point (2 conditions)
in the $(V_1,V_2)$-plane.

We now discuss all  $\vec{n}$ with $C_{4h}$ symmetry. 
Like $\vec{v}$, they are fully characterized by 
two values, $N_1$ and $N_2$, the total density at points 1--8 and
9--12. $n_{13} = 2-N_1-N_2$. 
$N_1$ and $N_2$ are confined within a stripe in the $(N_1,N_2)$ plane, with
an  ``inner'' boundary, with $n_{13}=1$,  reflecting the Pauli principle
($N_1 + N_2 < 1$ would imply $n_{13} > 1$), and an
``outer'' boundary, with $n_{13}=0$, reflecting particle conservation ($N_1 + N_2 \le 2$).

The manifold of allowed densities in the $(N_1,N_2)$ plane is subdivided into
different regions, corresponding to those of the $(V_1,V_2)$-plane.
All of the regions in the $(N_1,N_2)$ plane have $\rm D=2$. This is in line with
our general topological results for $g$-VR densities.

\begin{figure}
\unitlength1cm
\begin{picture}(5.0,8.5)
\put(-6.,-7.8){\makebox(5.0,8.5){
\includegraphics{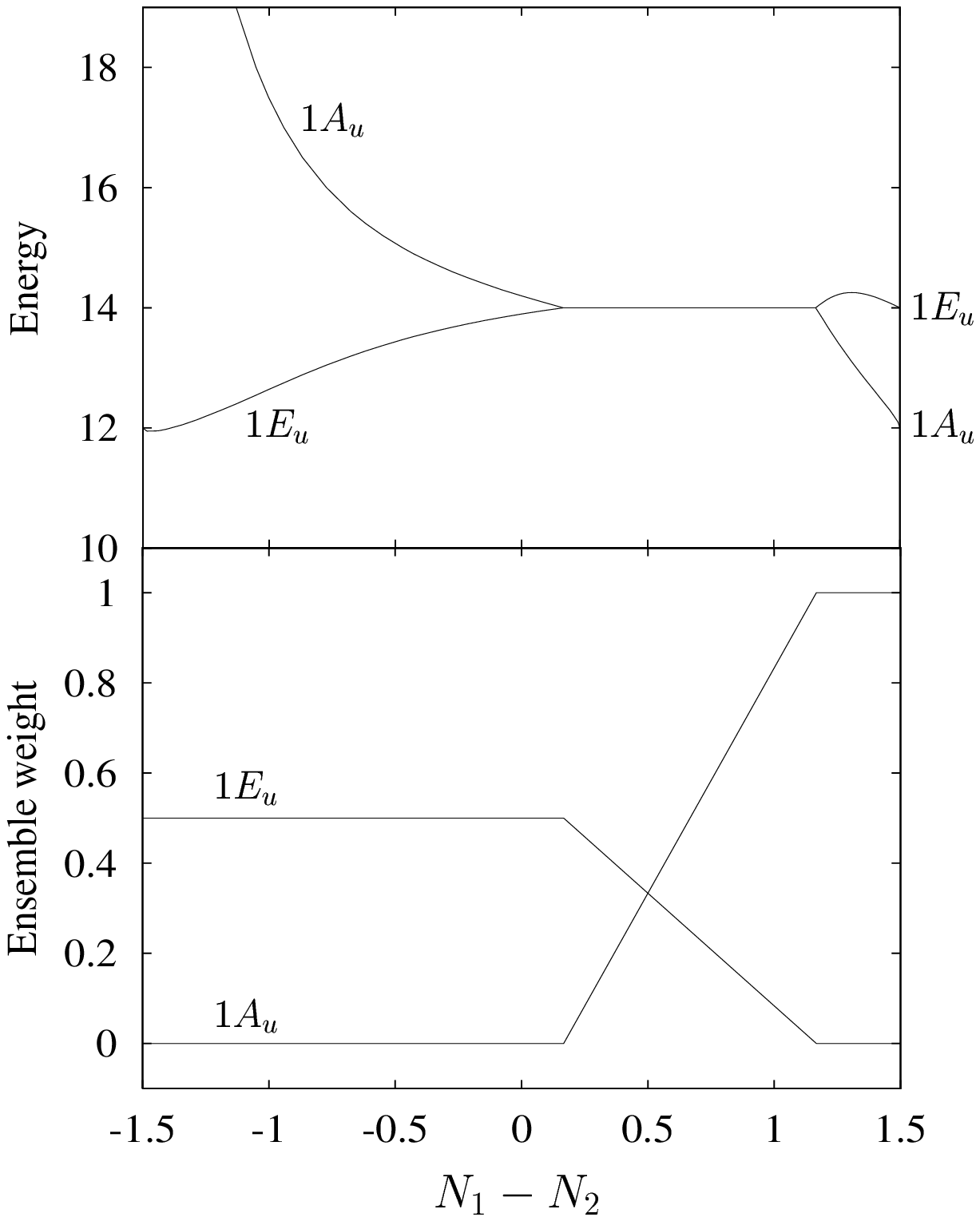}}}
\end{picture}
\caption{
Energies and ensemble weights of the next two levels 
above $1A_g$, along paths in the $(V_1,V_2)$
and $(N_1,N_2)$ planes (dotted lines in Fig. \ref{figure3}).
}
\label{figure4}
\end{figure}

Fig. \ref{figure4} illustrates the highest occupied state along
paths in the $(V_1,V_2)$ and $(N_1,N_2)$ planes (see dotted lines in Fig. \ref{figure3} top and bottom):
$1A_u$ with weight 1 (region 1a), $1E_u$ with weight 1
(equal-weight combination of $2p_x,2p_y$-like states; region 1c), or a linear
combination of the degenerate $1A_u$ and $1E_u$ (region 2a), with fractional weights.

{\bf Physical Example: the Be-series}.
Be has a nominal electronic configuration
$(1s)^2 (2s)^2$ and a term value $^1S$. Without interactions,
$2s$ and $2p$ are degenerate.
The Be-series consists of neutral Be ($Z = 4$)\cite{rvl}, and the
4-electron ions with $Z = 5,6,\ldots,\infty$. 
By re-scaling ($\tilde{\bf r} = Z{\bf r}/4$), this is equivalent to a series
of  Be-like atoms with $Z=4$ but with weakened interaction strength
$\tilde{e}^2= e^2(4/Z)$.  For $Z \to \infty$,
only the nearly degenerate configurations $(1s)^2 (2s)^2$ and $(1s)^2 (2p)^2$, 
$(2p)^2 = 3^{-1/2}\sum_{j=1}^3 (2p_j)^2$, need to be considered.
This leads to a $2\times 2$ secular equation all of whose MEs are exactly 
known \cite{shull}. The limiting density is explicitly $4$-VR:
\begin{equation}\label{Be_density}
n_0(\tilde{\bf r}) = 2\bigg[
\varphi_{1s}(\tilde{\bf r})^2
+ 0.949 \varphi_{2s}(\tilde{\bf r})^2
+ \frac{0.051}{3} \sum_{j=1}^3
\varphi_{2p_j}(\tilde{\bf r})^2 \bigg] ,
\end{equation}
where the $\varphi$'s are scaled hydrogenic wave functions (for $Z=4$).
The configurations
$(1s)^2 (2s)^2$ and $(1s)^2 (2p)^2$ have amplitudes $0.974$ and $0.225$.
(Unlike in CI expansions \cite{bunge}, KS densities consist
{\em strictly} of a {\em finite} number of orbital densities for {\em all} $Z$).

From our previous discussion of boundary surfaces, 
we observe that the density
(\ref{Be_density}) is {\em not} on a boundary,
since there is no unoccupied degenerate state.
Under these conditions, {\em any} density
$n({\bf r})$  sufficiently near $n_0({\bf r})$, in general without any symmetry, 
is also $4$-VR. In particular, there must be a finite range in $1/Z$ of
the Be-series where densities are $4$-VR. This is borne out by
numerical calculations \cite{savin}.

{\bf Density ensembles at $T=0$ and $0^+$}.
Let $\vec{v}^g$ be a potential with $g$ degenerate non-interacting, gapped ground
states, $\Phi_j$, generating the $\frac{1}{2}(g-1)(g+2)$-D set of ensemble densities
(\ref{nn}). Now introduce a weak perturbing potential $^1\vec{v}$ of order $\lambda$
which, in general, splits the degeneracy and leads to eigenstates and eigenvalues
$\Psi_j = \bar{R}^g \Phi_j + {\cal O}(\lambda)$; $E_j = E + {^1\!E_j}$,
where  $^1\!E_j = {\cal O}(\lambda)$,
and the corresponding densities are $\vec{n}_j$.

For a given low temperature, $\beta^{-1}\ll \Delta$, the canonical density
corresponding to $\vec{v}+{^1\vec{v}}$ is given by
\begin{equation}
\vec{n}' = {\cal Z}^{-1} \sum_{j=1}^g e^{-\beta \:{^1\!E_j}} \: \vec{n}(\bar{R}^g
\Phi_j) + {\cal O}(\lambda) + {\cal O}\left(e^{-\beta \Delta}\right),
\end{equation}
where ${\cal Z}= \sum_{j}^g e^{-\beta \:{^1\!E_j}}$. 
This density is equal, to leading order, to the density $\vec{n}$ of Eq. (\ref{nn}),
if $^1\vec{v}$ is chosen so that $\bar{R}^g\Phi_j = \Psi_j$ and 
${\cal Z}^{-1} e^{-\beta \:{^1\!E_j}} = w_j$.
As $\beta \Delta \to \infty$, 
$^1v \sim \beta^{-1} \to 0$ and $\vec{n}' \to \vec{n}$. This establishes the
correspondence between ensemble densities at $T=0$ and canonical densities
at low temperatures, $\beta \Delta \gg 1$.

{\bf Concluding remarks}.
The problem of $v$-represen\-tability \cite{Levy,Lieb} had cast a shadow of uncertainty
over DFT, which was very largely clarified in \cite{Chayes}. Here,
building on CCR, we clarify the topology of the $v({\bf r})$-$n({\bf r})$
mapping. The Be-series and a lattice model are used for illustration. 
The transition from a dense discrete lattice, $r_l$, to a continuous variable, ${\bf r}$,
does not appear to offer difficulties.
An open issue is extension to spin magnetism.

\begin{acknowledgments}
Support by  NSF  Grants No. DMR-96-30452 and DMR-99-76457 and discussions with A. Savin 
and C. Umrigar are gratefully acknowledged.
\end{acknowledgments}


\end{document}